\documentclass[10pt,twocolumn,letterpaper]{article}

\usepackage{iccv}
\usepackage{times}
\usepackage{epsfig}
\usepackage{graphicx}
\usepackage{amsmath}
\usepackage{amssymb}
\usepackage{booktabs}

\usepackage[accsupp]{axessibility}  

\usepackage[pagebackref=true,breaklinks=true,letterpaper=true,colorlinks,bookmarks=false]{hyperref}

\iccvfinalcopy 

\def\iccvPaperID{2104} 

\ificcvfinal\fi

\begin{document}

\title{Efficient Emotional Adaptation for Audio-Driven Talking-Head Generation}

\author{
Yuan Gan$^{1,2}$, \quad Zongxin Yang$^{1,2}$, \quad Xihang Yue$^{1,2}$, \quad Lingyun Sun$^{2}$,  \quad Yi Yang$^{1,2}$\thanks{Corresponding author.}\\
$^1$ReLER, CCAI, Zhejiang University, China \\
$^2$College of Computer Science and Technology, Zhejiang University, China
\\ 
}

\maketitle
\ificcvfinal\thispagestyle{empty}\fi

\begin{abstract}

Audio-driven talking-head synthesis is a popular research topic for virtual human-related applications. However, the inflexibility and inefficiency of existing methods, which necessitate expensive end-to-end training to transfer emotions from guidance videos to talking-head predictions, are significant limitations. In this work, we propose the Emotional Adaptation for Audio-driven Talking-head (EAT) method, which transforms emotion-agnostic talking-head models into emotion-controllable ones in a cost-effective and efficient manner through parameter-efficient adaptations. Our approach utilizes a pretrained emotion-agnostic talking-head transformer and introduces three lightweight adaptations (the Deep Emotional Prompts, Emotional Deformation Network, and Emotional Adaptation Module) from different perspectives to enable precise and realistic emotion controls. Our experiments demonstrate that our approach achieves state-of-the-art performance on widely-used benchmarks, including LRW and MEAD. Additionally, our parameter-efficient adaptations exhibit remarkable generalization ability, even in scenarios where emotional training videos are scarce or nonexistent. Project website: \href{https://yuangan.github.io/eat/}{https://yuangan.github.io/eat/}

\end{abstract}

\section{Introduction}

Recently, there has been increasing attention on synthesizing realistic talking heads due to their wide-ranging applications in industry, such as digital human animation~\cite{guo2021ad, ji_eamm:_2022, wang2022one}, visual dubbing~\cite{prajwal2020lip}, and video content creation~\cite{song2021everything}.
Audio-driven talking-head generation aims to produce realistic talking-head videos synchronized with speech. However, unlike speech, humans convey intentions through emotional expressions. Therefore, generating emotional talking heads is important to improve the fidelity of talking heads for real-world applications. To address this open problem, various forms of knowledge (such as human head models, emotions, audio, and vision) must be considered in constructing multi-knowledge representations~\cite{mkr}.

\begin{figure}
  \centering
  \includegraphics[width=0.48\textwidth]{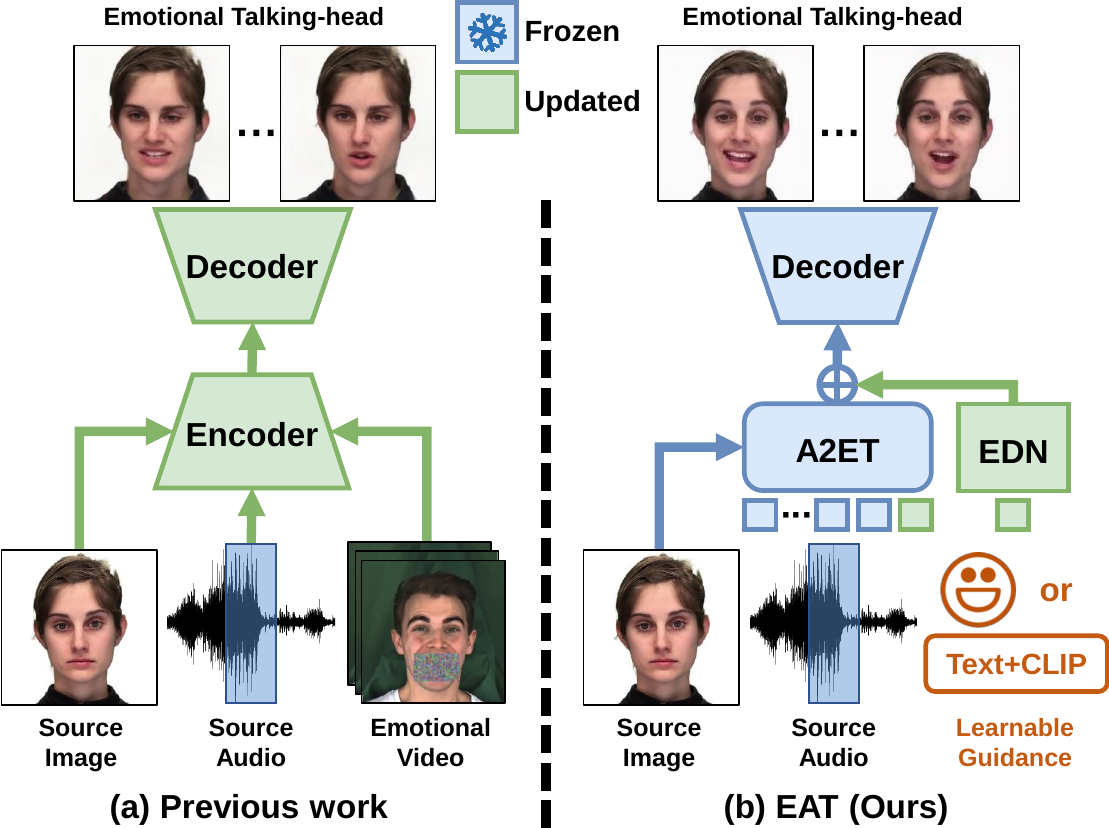}
  \caption{\textbf{Efficient emotional talking-head generation.} (a) Previous work trains or finetunes the whole network with augmented emotional driving videos. (b) Our EAT transforms emotion-agnostic talking-head
models into emotion-controllable ones through flexible guidance, including emotional prompts or text-guided CLIP~\cite{radford2021learning} supervision, by lightweight adaptations.}
  \label{fig:fig_first} 
  \vspace{-0.1cm}
\end{figure}

Previous one-shot talking-head generation methods~\cite{zhou2020makelttalk, prajwal2020lip, wang2022one} focus on achieving audio-visual synchronization for emotion-agnostic talking-heads, which is a special case of realistic talking-heads. More recent works~\cite{ji2021audio, ji_eamm:_2022, liang2022expressive} pay attention to generating emotion-aware talking-heads. GC-AVT~\cite{liang2022expressive} and EAMM~\cite{ji_eamm:_2022} are two methods that generate emotional videos using driven emotional videos and pose-guiding videos. GC-AVT~\cite{liang2022expressive} achieves explicit control over the expression, speech content, and pose of talking heads through granular pre-processing design. EAMM~\cite{ji_eamm:_2022} synthesizes one-shot emotional talking-heads by adding augmented emotional source videos. Since the driven videos can introduce semantic ambiguity to mouth shapes, GC-AVT replaces the mouth part with the neighbor frame, while EAMM ignores the mouth part of driven emotional videos by using data augmentation. Additionally, these methods require training or finetuning the entire network at high costs for emotional talking-heads generation.

Although emotion-aware methods have made progress in the one-shot talking-head generation, they lack in-depth thinking in two key aspects. \textbf{(1) Architecture efficiency.} As a sub-task of talking-head generation, it is parameter-inefficient to train or finetune the entire emotional talking-head generation network. Furthermore, since large-scale emotion-agnostic talking-head data is more readily available than emotional data,
it is worthwhile to consider how to efficiently reuse the knowledge learned from emotion-agnostic data. \textbf{(2) Guidance flexibility.} Previous methods prefer to transfer the driving videos to the target talking heads rather than directly learning the emotional representation. In practice, finding appropriate driven emotional videos requires taking into account factors such as resolution, occlusion, and even the length of the driven emotional videos and audio. Furthermore, prior research has neglected to consider lip shapes, which can result in unrealistic emotional expressions. For instance, according to FACS~\cite{ekman1978facial, keltner2019emotional}, depressed lip corners are one of the key components of the sad expression.

To address the above limitations, a desirable approach should enable an efficient and flexible transfer of pretrained talking-head models to emotional talking-head generation tasks with lightweight emotional guidance, as illustrated in Fig.~\ref{fig:fig_first}. There are two key advantages.
Firstly, with the reused knowledge, we can readily and effortlessly apply the talking-head model to emotional talking-head generation tasks. Secondly, obtaining lightweight guidance is simpler and more adaptable in practical scenarios, such as text-guided zero-shot expression editing.

To realize the aforementioned paradigm, we propose an efficient Emotional Adaptation framework for audio-driven Talking-head (EAT) generation, which involves two stages. In the first stage, we enhance the unsupervised 3D latent keypoints representation~\cite{wang2021one} to capture emotional expressions.  Then, we introduce the Audio-to-Expression Transformer (A2ET), which learns to map audio to enhanced 3D latent keypoints using large-scale talking-head datasets. In the second stage, we propose learnable guidance and adaptation modules for steering emotional expression generation. These include Deep Emotional Prompts for parameter-efficient emotional adaptation, a lightweight Emotional Deformation Network (EDN) for learning the emotional deformation of facial latent representation, and a plug-and-play Emotional Adaptation Module (EAM) for enhancing visual quality. Our approach enables rapid transfer of traditional talking-head models to emotional generation tasks with high-quality results and supports zero-shot expression editing with image-text models~\cite{radford2021learning}.

We conduct extensive experiments 
to assess the effectiveness of EAT on emotional talking-head generation. Compared to baseline competitors, EAT achieves superior performance without guiding emotional videos. Moreover, based on the pretrained talking-head model, we can attain state-of-the-art (SOTA) performance in 2 hours with only 25\% training data.
The results indicate that our method is capable of generating more realistic talking-head videos. And with only text descriptions of emotions, we can achieve zero-shot talking-head editing.

In summary, the main contributions of our work are listed below:
\begin{itemize}
    \item Our study introduces a new two-stage paradigm, called EAT, for addressing emotional talking-head tasks. Our experiments demonstrate that this paradigm outperforms previous methods with respect to both emotion manipulation and video quality in one-shot talking-head generation tasks.
    \item Our proposed architecture includes deep emotional prompts, an emotional deformation network, and an emotional adaptation module. This design enables the efficient transfer from generating talking heads without emotional expression to generating talking heads with emotional expression.
    \item To the best of our knowledge, our study is the first to introduce flexible guidance for talking-head adaptation. By utilizing image-text models, we can achieve zero-shot expression editing of talking-head videos, surpassing the capabilities of previous methods.

\end{itemize}

\section{Related Work}
\label{sec:related}
Previous works have achieved great performance in audio-visual synchronization.
However, there remain some challenges such as efficient knowledge transfer based on a large-scale pretrained talking-head model.

\noindent\textbf{Audio-driven Talking Head Generation.}
Audio-driven talking-head generation~\cite{chung2017you,song2018talking,zhou2019talking,chen2019hierarchical,zhou2020makelttalk,guo2021ad,wang2022one,zhou2021pose}
based on deep learning has attracted lots of attention in recent years. 
Chen~\emph{et al.}~\cite{chen2019hierarchical} design a two-stage structure that leverages facial landmarks as an intermediate representation
MakeItTalk ~\cite{zhou2020makelttalk} generates one-shot talking heads based on disentangled speech and speaker. PC-AVS~\cite{zhou2021pose} generates arbitrary talking heads with pose control. 

Following the groundbreaking work of transformers~\cite{vaswani2017attention} in NLP~\cite{devlin2018bert,radford2019language}, recent works have achieved remarkable progress in video-related tasks, including action recognition~\cite{arnab2021vivit,fan2021multiscale,liu2022video}, video segmentation~\cite{yang2021associating,yang2022decoupling,zhu2022instance,pan2022n}, cross-model understanding~\cite{li2023seg,samtrack}, etc. In the talking-head field,
AVCT~\cite{wang2022one} designs an audio-visual correlation transformer for generating talking-head videos.
In this work, we enhanced 3D latent keypoints and applied transformers to generate more realistic talking heads.
\begin{figure*}[ht!]
  \centering
  \includegraphics[width=1.00\textwidth]{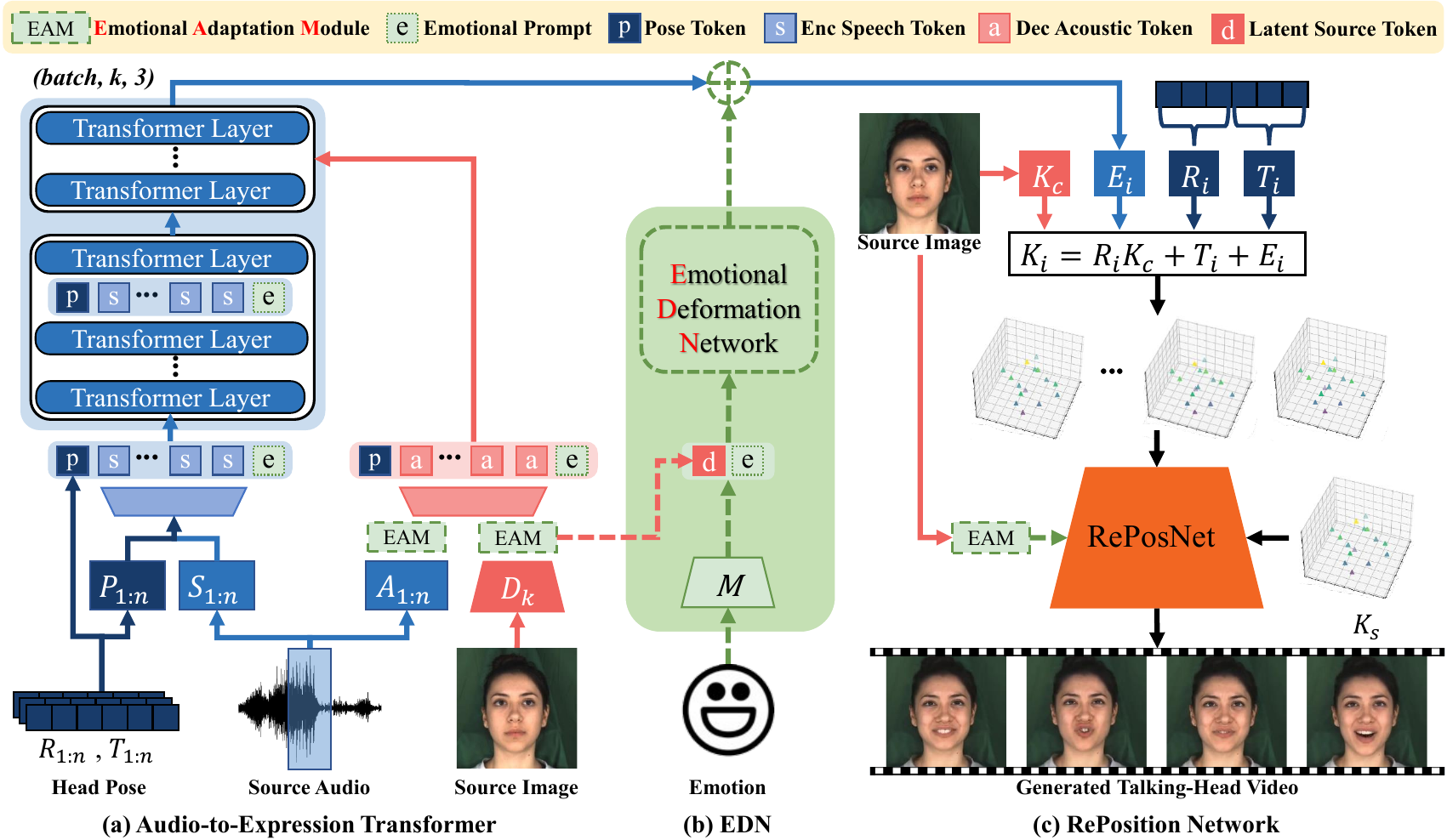}

  \caption{\textbf{Overview of EAT model.} (a) In the first stage, the Audio-to-Expression Transformer (A2ET) transfers latent source image representation, source audio and head pose sequences to 3D expression deformation. (b) In the second stage, the emotional guidance is injected into A2ET, Emotional Deformation Network (EDN) and Emotional Adaptation Module (EAM) for emotional talking-head generation, presented in dashed lines. (c) The RePos-Net takes the 3D source keypoints $K_{s}$ and driven keypoints $K_{i}$ to generate frames.}  
  \label{fig:arch}
\vspace{-0.1cm}
\end{figure*}

\noindent\textbf{Emotion-aware Talking Head Generation.} 
Emotional talking-head generation has been studied recently for realistic talking-head generation.
Pumarola~\emph{et al.}~\cite{pumarola2018ganimation} introduce an unsupervised framework to generate facial videos with a specific expression. EVP~\cite{ji2021audio} proposes emotional video portraits to produce more vivid results based on source video. However, the one-shot emotional talking-head generation has emerged recently. Sinha~\cite{sinha2022emotion} generates pose-fixed emotional talking-heads with graph convolution. GC-AVT~\cite{liang2022expressive} train an emotion and pose controllable model with a granular pre-processing design. EAMM~\cite{ji_eamm:_2022} 
synthesizes one-shot emotional talking heads with augmented emotional source videos. In our work, we achieve emotion control with efficient adaptation based on a pretrained talking-head model.

\noindent\textbf{Efficient Finetuning.}
Efficient finetuing has been studied for knowledge transfer and many techniques have been proposed including residual adapter~\cite{rebuffi2017learning}, bias tuning~\cite{cai2020tinytl} and side tuning~\cite{zhang2020side}. Recently, prompt ~\cite{liu_pre-train_2022} has attracted more attention in vision task.
Inspired by prompt tuning in the language model, prompt tuning recently has been proposed in various visual tasks~\cite{radford2021learning, jia2021scaling, zhou2022coop, jia2022vpt, zhou2022cocoop, du2022learning, li2022bridge} for effectiveness and efficiency. 
CoOp~\cite{zhou2022coop} and VPT~\cite{jia2022vpt} utilize learnable prompt vectors and achieve better performance. Moreover, to increase generalizability, CoCoOp~\cite{zhou2022cocoop} designed a lightweight network to learn prompt vectors for each image. In our work, we introduce deep emotional prompts, an emotional deformation network and emotional adaptation modules to achieve efficient and effective emotion-related knowledge transfer.

\section{Method}
To avoid the expensive end-to-end training and finetuning of previous methods~\cite{ji_eamm:_2022, liang2022expressive}, we propose a two-stage paradigm, the efficient Emotional Adaptation
for audio-driven Talking-head (EAT) generation method. Firstly, we introduce the enhanced 3D latent representations and the emotion-agnostic talking-head pretraining using the Audio-to-Expression Transformer (A2ET). (Sec.~\ref{sec:3.1}) Secondly, we present a parameter-efficient emotional adaptation approach to quickly adapt the pretrained talking-head model for emotional talking-head tasks. This approach includes deep emotional prompts, an Emotional Deformation Network (EDN), and an Emotional Adaptation Module (EAM). (Sec.~\ref{sec:3.2}) At last, we introduce our training objectives in detail. (Sec.~\ref{sec:3.3})

 \begin{figure}
  \centering
  \includegraphics[width=0.48\textwidth]{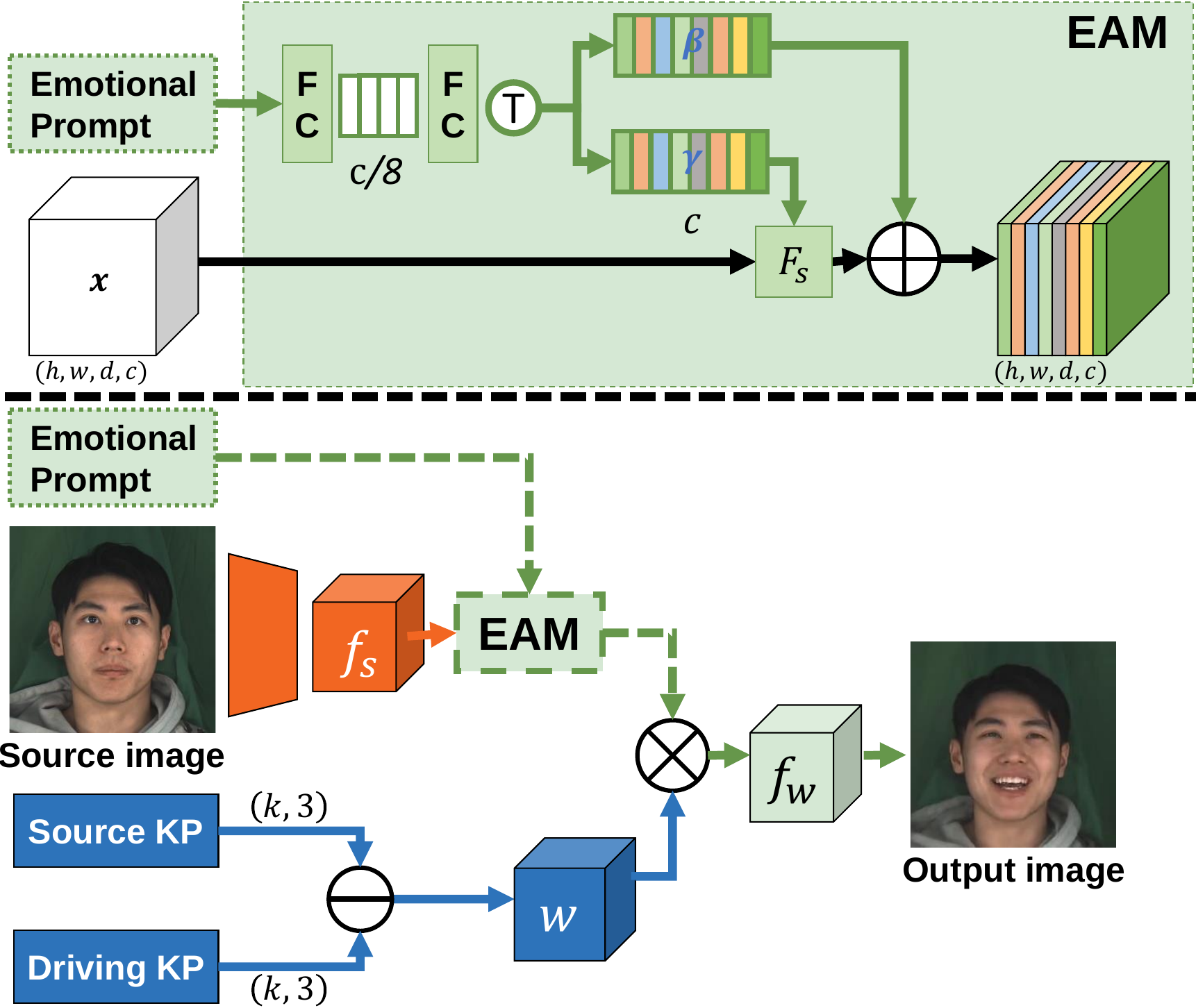}
  \caption{\textbf{The architecture of RePos-Net and EAM.} RePos-Net extracts 3D appearance features $f_{s}$ from the source image. Given the 3D source keypoints and driving keypoints, RePos-Net predicts the 3D flow warp matrix $w$ to transform the 3D feature $f_{s}$ and generates the output frame. EAM transfers emotional guidance to emotion-conditioned features with the learned $\gamma$ and $\beta$.}
  \label{fig:RePosnet} 
  \vspace{-0.1cm}
\end{figure}

\subsection{Emotion-agnostic Pretraining}
\label{sec:3.1}
Compared to inefficient end-to-end training for emotional talking-head~\cite{liang2022expressive, ji_eamm:_2022}, we explore leveraging the knowledge gained from an adaptable model that has been pre-trained on emotion-agnostic datasets. The model allows for rapid adaptations to downstream tasks, including emotional talking-head generation.
To achieve such an emotion-agnostic pertaining model, we firstly enhance the 3D latent representation~\cite{wang2021one} to capture the subtle expression better. Then an Audio-to-Expression Transformer (A2ET) is introduced to capture the temporal context of audio clips, generate audio-visual synchronized 3D latent expression sequences for talking-head generation, and support efficient emotional adaptations (Sec.~\ref{sec:3.3}).

\subsubsection{Enhanced Latent Representation.}
\label{sec:3.1.1}
Given a talking-head frame $i$, the 3D latent keypoints $K_i$, which are learned unsupervisedly, are composed of four components: identity-specific canonical keypoints $K_c$, rotation matrix $R_i$, translation $T_i$, and expression deformation $E_i$. These components are then combined with the following equation:
\begin{equation}
K_{i} = R_{i}K_{c}+T_{i}+E_{i}. \label{eq:ki}
\end{equation}

Based on 3D latent keypoints, \textit{RePosition Network} (RePos-Net)~\cite{wang2021one} can transfer facial expressions from one person to another, as illustrated in Fig.~\ref{fig:RePosnet}. However, we observe that this kind of transfer fails to account for other facial expression elements, such as eyebrows, lip corners, etc. Therefore, we enhance the latent representation for more realistic talking-head generation with these modifications:
\begin{itemize}
\item We remove the deformation prior loss in OSFV~\cite{wang2021one} which penalizes the magnitude of the keypoints deformation. This allows our latent keypoints to capture more subtle changes in facial expressions.
\item We use the MEAD dataset~\cite{wang2020mead} to acquire labeled and paired face data from neutral and emotional videos of the same identity. This helps the network learn more expressive faces from expression changes. 
\item To avoid the influence of expression-irrelevant background, we only compute losses on the facial part. And we augment our paired data with the affectnet~\cite{8013713} dataset to improve generalizability. 
\end{itemize}

These modifications enhance the representation ability of the learned 3D latent keypoints, which are the objective of our A2ET model.

\subsubsection{Audio-to-Expression Transformer.}
\label{sec:3.1.2}
Since the 3D latent keypoints are specific to the source identity and more complex than 2D latent keypoints~\cite{siarohin2019first, wang2022one, ji_eamm:_2022}, predicting the 3D keypoints sequence directly is a challenging task. We notice that the facial expressions are mainly represented by the expression deformation $E_{i}$ in 3D latent keypoints. Thus, the purpose of A2ET is to learn the audio-visual synchronized expression deformation, which is composed of audio-visual feature extraction and expression deformation prediction.

\noindent\textbf{Audio-visual Feature Extraction.} 
Previous work~\cite{wang2022one} generates emotion-agnostic talking heads with transformer and phonemes. However, training transformer requires a large dataset, and phoneme extraction is challenging in noisy or accented speech. To address these limitations, we train our A2ET model on a large dataset Voxceleb2~\cite{chung_voxceleb2:_2018} and extract speech features $S_{1:n}$ and acoustic features $A_{1:n}$ as inputs. The audio semantic features $S_{1:n}$ are extracted via a speech recognition model~\cite{amodei2016deep} from the MFCC features. To derive acoustic features $A_{1:n}$, we design an audio encoder to encode the mel spectrogram extracted with 80 mel bins and 1025 frequency bins. 

As shown in Fig.~\ref{fig:arch}(a), given frame $i$, we extract its semantic context features from $2w+1$ audio frames. Initially, speech features $S_{i-w:i+w}$ and head pose features $P_{i-w:i+w}$ are converted into speech tokens. The 6DoF of frame $i$ is encoded into a pose token $p$. The A2ET encoder takes these tokens as input.
Subsequently, to capture subtle mouth movements, we encode acoustic features $A_{i-w:i+w}$ and the latent source image representation with an audio encoder~\cite{kum_joint_2019} and keypoint detector $D_{k}$. These representations are fused to obtain acoustic tokens that are used by the A2ET decoder to output the feature of $2w+1$ tokens.

\begin{figure*}
  \centering
  \includegraphics[width=1.0\textwidth]{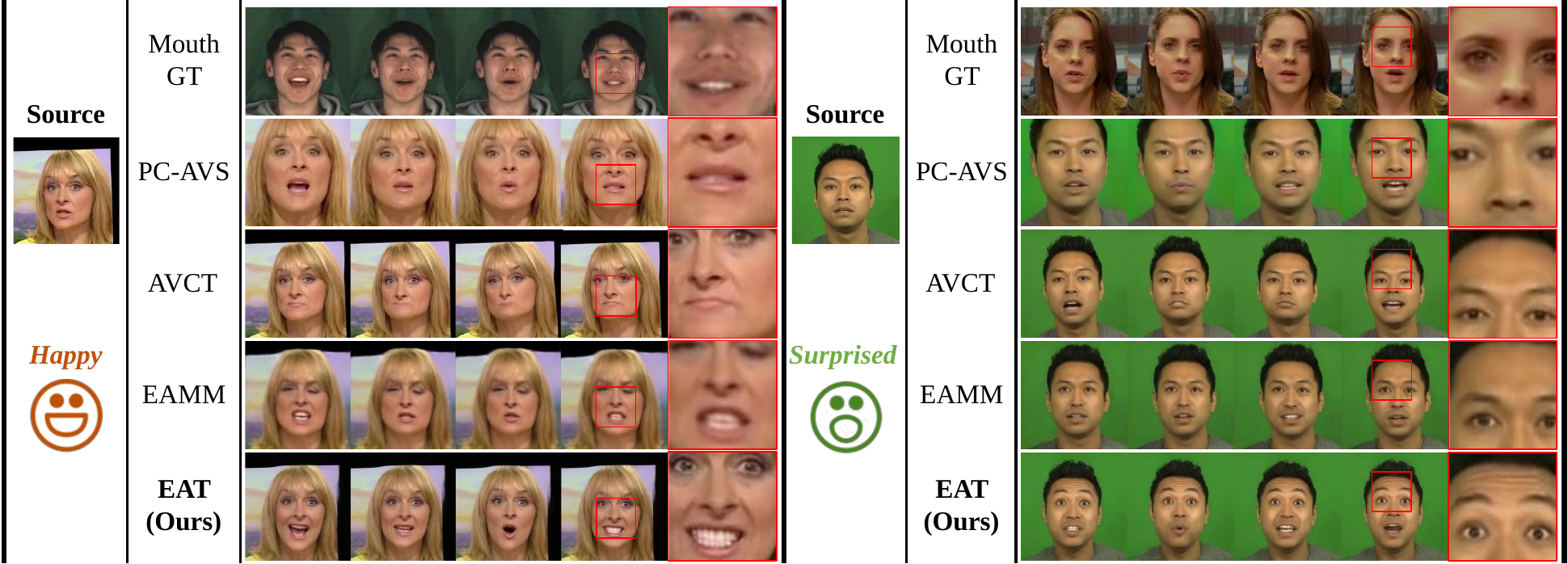}
  \caption{\textbf{The qualitative results of the one-shot emotional talking-head generation cases.} Results of \textit{happy} and \textit{surprised} are on the left and right. The top row shows the identity and driving emotion. The second row shows the content ground-truth frames. The left source face from LRW~\cite{chung2016lip}. The right face from CREMA-D ~\cite{cao2014crema}.}
  \label{fig:qualitative}
  \vspace{-0.1cm}
\end{figure*}

\noindent\textbf{Expression Deformation Prediction.} The expression deformation $E_{i}$, composed of $k$ 3D offsets, can be predicted from the feature of central frame $i$, but optimizing the 3D expression motions leads to convergence issues. We observed that the 3D keypoints learned by self-supervision exhibit inherent inter-dependencies, with only a few keypoints influencing facial expressions. To address this, the principal component analysis (PCA) of $E_{i}$ is adopted to reduce the dimensionality and eliminate nonessential information.  
Then we can predict the 3D expression deformation from the audio features.

\subsection{Efficient Emotional Adaptation}
\label{sec:3.2}
Traditional audio-driven talking-head methods~\cite{chen2019hierarchical,zhou2020makelttalk,guo2021ad,wang2022one,zhou2021pose} have made impressive advancements in emotion-agnostic talking-head generation tasks. However, to achieve realistic talking heads, emotional expression is crucial. Hence, we present a novel approach that includes three parameter-efficient modules for swift emotional adaptation from emotion-agnostic models. These modules comprise Deep Emotional Prompts, the Emotional Deformation Network (EDN), and the Emotional Adaptation Module (EAM), specifically designed to enable efficient emotional adaptation of pre-trained A2ET. Our approach allows for lightweight adaptations, which offer flexibility in guiding downstream tasks, such as zero-shot expression editing.

\noindent\textbf{Emotional Guidance.} One straightforward idea is to generate emotional talking heads using learnable guidance conditioned on emotions. 
We posit that each emotional type belongs to a distinct sub-domain in latent space. As shown in Fig.~\ref{fig:arch} (b), a mapping network $M$ is adopted to extract emotion-conditional guidance with a latent code $z \in \mathcal{U}^{16}$. the latent code is sampled from a Gaussian distribution, which is commonly used in generative models~\cite{Kingma2014, choi2020stargan}. This emotional guidance is applied to steer the generation of emotional expressions.

\noindent\textbf{Deep Emotional Prompt.}
To achieve parameter-efficient emotional adaptation, we include the guidance as an additional input token of the A2ET transformer layer, as shown in Fig.~\ref{fig:arch}(a).
We separately introduce shallow and deep emotional prompts into the A2ET transformer architecture, with the shallow prompt added to the first layer and the deep prompt added to every layer thereafter. Our results in Table~\ref{tab:prompt} demonstrate that the deep prompt leads to better emotional expression transfer compared to the shallow prompt. However, we also observe that incorporating emotional prompts can have a detrimental effect on audio-visual synchronization. Generating emotional expressions using fixed transformer weights while ensuring audio-visual synchronization may pose a challenge for the prompts.

\begin{table*}
\setlength{\tabcolsep}{5pt}
\begin{center}
\small
\begin{tabular}{l| c c c c | c c c c c}
\toprule
 & & &\hspace{-4.2em}LRW~\cite{chung2016lip}     & & & &MEAD~\cite{wang2020mead} & &
\\
              & PSNR/SSIM$\uparrow$ &FID$\downarrow$ & SyncNet$\uparrow$ & M/F-LMD$\downarrow$ & PSNR/SSIM$\uparrow$ &FID$\downarrow$ & SyncNet$\uparrow$ & M/F-LMD$\downarrow$ & Acc$_{emo}$$\uparrow$ \\
\midrule
ATVG~\cite{chen2019hierarchical}    &18.40/0.64 &51.56 &2.73 &2.69/3.31 &17.64/0.56 &99.42 &1.80 &2.77/3.74 &17.36\\
Wav2Lip~\cite{prajwal2020lip}       &22.80/0.73 &7.44  &\textcolor{red}{7.59}  &\textcolor{blue}{1.58}/2.47 &19.12/0.57 &67.49 &\textcolor{red}{8.97} &3.11/3.71 &17.87\\
MakeItTalk~\cite{zhou2020makelttalk}    &21.67/0.69 &3.37  &3.28 &2.16/2.99 &18.79/0.55 &51.88 &5.28 &3.61/4.00 &15.23\\
AVCT~\cite{wang2022one}          &21.72/0.68 &\textcolor{blue}{2.01}  &4.63 &2.55/3.23 &18.43/0.54 &39.18 &6.02 &3.82/4.33 &15.64\\
PC-AVS~\cite{zhou2021pose}        &23.32/0.72 &4.64  &\textcolor{blue}{7.36} &\textcolor{red}{1.54}/\textcolor{blue}{2.11} &\textcolor{blue}{20.60}/0.61 &53.04 &\textcolor{blue}{8.60} &2.66/2.70 &11.88\\
EAMM~\cite{ji_eamm:_2022}          &22.34/0.71 &6.44  &4.67 &1.81/2.37 &20.55/\textcolor{blue}{0.66} &\textcolor{blue}{22.38} &6.62 &\textcolor{red}{2.19}/\textcolor{blue}{2.55} &\textcolor{blue}{49.85}\\
\midrule
\textbf{Pretrain (Ours)}    &\textcolor{blue}{23.97}/\textcolor{blue}{0.76} &\textcolor{red}{1.89}  &6.30 &1.95/2.12 &20.32/0.61 &26.71 &8.09 &2.83/2.99 &25.18\\
\textbf{EAT (Ours)}     &\textcolor{red}{24.11}/\textcolor{red}{0.77} &3.52  &6.22 &1.79/\textcolor{red}{2.08} &\textcolor{red}{21.75}/\textcolor{red}{0.68}  &\textcolor{red}{19.69} &8.28 &\textcolor{blue}{2.25}/\textcolor{red}{2.47} &\textcolor{red}{75.43}\\
\midrule
Ground Truth  &  $\infty$ /1.00 &0     &7.06 &0.00/0.00 &  $\infty$  /1.00 &0     &7.76  &0.00/0.00 &84.37\\
\bottomrule
\end{tabular}

\end{center}
   \vspace{-0.15cm}
\caption{\small{\textbf{Quantitative comparisons with state-of-the-art methods on LRW~\cite{chung2016lip} and MEAD~\cite{wang2020mead}.} We present the results of pretrained A2ET and full EAT model on both LRW and MEAD.  M/F-LMD denotes the landmark distance of mouth and face. "$\uparrow$": higher is better. "$\downarrow$": lower is better. \textcolor{red}{Red}: the 1st score. \textcolor{blue}{Blue}: the 2nd score.}
}
  \vspace{-0.25cm}
\label{tab:tab_lrw}
\end{table*}

\noindent\textbf{Emotional Deformation Network.} We observe that the decoupled 3D implicit representations in Eq.~\ref{eq:ki} exhibit linear additivity. Furthermore, emotional talking heads show emotional deformation that is not present in traditional talking-heads. To complement $E_{i}$, one intuitive approach is to include an emotional expression deformation term:
\begin{equation}
  E^{\prime}_{i} = E_{i} + \Delta E_{i}, \label{eq:eee}
\end{equation}
where $E^{\prime}_{i}$ represents emotional expression deformation, $E_{i}$ represents speech-related expression deformation predicted by A2ET, and $\Delta E_{i}$ represents emotion-related expression deformation. To predict $\Delta E_{i}$, we design a sub-network called the Emotional Deformation Network (EDN), which is depicted in Fig.~\ref{fig:arch}(b). EDN utilizes the A2ET encoder architecture to predict $\Delta E_{i}$ with emotional guidance and the source latent representation token. To accelerate adaptation, we initialized EDN with the pretrained A2ET encoder.
To update $E_{i}$ with $E^{\prime}_{i}$, we can get emotional 3D latent keypoints using Eq.~\ref{eq:ki}. 

\noindent\textbf{Emotional Adaptation Module.} To enhance the visual quality, we have designed a lightweight, plug-and-play adaptation module called the Emotional Adaptation Module (EAM) that can generate emotion-conditioned features. As shown in Fig.~\ref{fig:RePosnet}, the module takes in the guidance embedding $e$ and processes it through two fully connected (FC) layers to obtain a set of channel weights $\gamma$ and bias $\beta$. 
And we use the tanh activation function to confine the $\gamma$ and $\beta$ values to the range [-1, 1]:
\begin{equation}
  \gamma, \beta = \tanh(\text{FC}(\text{ReLU}(\text{FC}(e)))).
\end{equation}
Once we have obtained $\gamma$ and $\beta$, we can input the feature $x$ to obtain the emotional feature, which is calculated using the following equation:
\begin{equation}
  EAM(x) = F_{s}(1+\gamma, x) + \beta, \label{eq:eam}
\end{equation}
where $F_s$ denotes the channel-wise multiplication. As shown in Figure~\ref{fig:arch}, the EAM can be inserted into RePos-Net, as well as the audio and image feature extractors.

\begin{table}[t]
  \begin{center}
      \setlength{\tabcolsep}{6pt}
      \small
          \begin{tabular}{l | c c c c c}
                \toprule
              Method & Wav2Lip & PC-AVS & EAMM & EAT & GT  \\
                \midrule
              Lip-sync  & 3.86 & 3.90 & 3.64 & \bf{3.99} & 4.59 \\
              Quality & 2.69 & 3.19 & 2.89 & \bf{3.35} & 4.59 \\
              Acc$_{emo}$& 13\% & 20\% & 35\% & \bf{50\%} & 66\% \\
                \bottomrule
          \end{tabular}
  \end{center}
   \vspace{-0.05cm}
    \caption{\textbf{User Study on CREMA-D and LRW.} Lip-sync and Quality represent the audio-visual synchronization and visual quality. Emotion classification accuracy (Acc$_{emo}$) evaluates the effectiveness of methods in producing emotional expressions.}
  \label{tab:userstudy}
   \vspace{-0.2cm}
\end{table}

\noindent\textbf{Zero-shot Expression Editing.} Owing to the quick adaptation capabilities of our EAT, we can achieve zero-shot text-guided expression editing of talking heads by distilling knowledge from CLIP~\cite{radford2021learning}, a large-scale vision-language pertaining model. This unique ability sets our work apart from the latest research~\cite{ji_eamm:_2022}, as it eliminates the need for emotional training data and enables generalization to applications that require rare expressions.

Specifically, our goal is to employ CLIP loss to learn emotional guidance correlated with text-described expressions.
To achieve this, we extract the head poses, source audio and the first frame from the target video as the input. Besides, a target expression description is taken for finetuning. Utilizing the refined EAT model and our training loss, we add an \textit{additional} CLIP loss~\cite{patashnik2021styleclip} to finetune the mapping network and EAM module only. In detail, we extract image embeddings from the predicted talking face using the image encoder of CLIP, and text embeddings from the description using its text encoder. We then iteratively optimize the distance between the image and text embeddings to align the generated talking face with the input text.

\subsection{Training Objectives}
\label{sec:3.3}
For supervised learning, the loss is calculated as follows:
\begin{equation}
\mathcal{L} =  \lambda_{lat}\mathcal{L}_{lat} + \lambda_{sync}\mathcal{L}_{sync} + \lambda_{rec}\mathcal{L}_{rec},
\end{equation}
where $\lambda_{lat}$, $\lambda_{sync}$ and $\lambda_{rec}$ are hyper-parameters that re-weight the corresponding term. As for zero-shot editing, we replace the $\lambda_{rec}\mathcal{L}_{rec}$ with the CLIP loss since there is no ground-truth video. 
In the following, we will discuss each training loss in detail.

\noindent\textbf{Latent Loss.} Latent loss is applied to optimize the predicted latent keypoints: 
\begin{equation}
\begin{aligned}
\mathcal{L}_{lat} = \frac{1}{N} \sum_{i=1}^N (
    \big\|  PE_{i} - \hat{PE}_{i}   \big\|_2^2 +
    \big\|  K_{i} - \hat{K}_{i}   \big\|_2^2), \label{eq:lat}
\end{aligned}
\end{equation}
where $N$ denotes the frame length of a sampled audio clip in each batch. $PE_{i}$ represents the predicted PCA of expression deformation in $i$ frame. $K_{i}$ is the transferred 3D latent keypoints according to Eq.~\ref{eq:ki}. $\hat{PE}_{i}$ and $\hat{K}_{i}$ are the corresponding ground-truth of the frame $i$. As the emotional expression deformation $\Delta E_{i}$ is not included in $P_{i}$, we only use the loss of 3D keypoints in Eq.~\ref{eq:lat} while training EDN.

\noindent\textbf{Sync Loss.} The synchronization loss is introduced in Wav2Lip~\cite{prajwal2020lip}. Based on the structure of SyncNet~\cite{chung2016out}, we train an expert to discriminate the audio-visual synchronization in neutral and emotional datasets. For a sampled audio clip in each batch, we computed the synchronization loss of the generated video using the following equation:
\begin{equation}
  \begin{aligned}
    \mathcal{L}_{\mathrm{sync}} = - \log(\dfrac{v \cdot s}{max(\lVert v \rVert_{2} \cdot \lVert s \rVert_{2}, \epsilon)}). \label{eq:sync}
  \end{aligned}
  \end{equation}
The input speech embedding $s$ and the generated video embedding $v$ are extracted by the speech encoder and image encoder in SyncNet, respectively.

\noindent\textbf{Reconstruction Loss.} To improve the expression generation, we employ $\mathcal{L}_1$ reconstruction loss only in the facial region. Additionally, to generate sharper frames, we apply perceptual loss~\cite{johnson2016perceptual} to the entire frames using a pretrained VGG19 model.

\noindent\textbf{CLIP Loss.} The CLIP loss calculates the similarity between the embeddings of the generated face and the text description with a pretrained CLIP model. Specifically, the CLIP loss is calculated by taking the cosine similarity between the normalized embedding of the image and text.

\section{Experiments}
\subsection{Experimental Setup}
\noindent\textbf{Implementation details.}
The videos are sampled to 25 FPS and the audio sample rate is 16KHz. The videos are cropped and resized to 256$\times$256. 
To synchronize the audio features and the video, We extract the mel-spectrogram~\cite{logan2000mel} by configuring the window length and hop length to 640.
The number of keypoints $k$ used in EAT is 15.
The mapping network of emotional prompts consists of shared four MLP layers and unshared four MLP layers for every kind of emotion.
We enhance the 3D latent keypoints for 48 hours and pretrain the A2ET with enhanced latent keypoints for 48 hours. Then we finetune the EAT architecture for 6 hours.
Our work is based on 4 NVIDIA 3090 GPUs.

\noindent\textbf{Datasets.}
 The training dataset consists of videos from VoxCeleb2~\cite{chung_voxceleb2:_2018} and MEAD~\cite{wang2020mead}. MEAD is a high-quality emotional talking-head video set with 8 kinds of emotions. To ensure fair comparisons, we split the MEAD dataset into training and testing sets based on identity, using the same test identities as EAMM~\cite{ji_eamm:_2022}. To learn large head pose changes, we select about 8,000 emotional videos from the VoxCeleb2~\cite{chung_voxceleb2:_2018} dataset with Emotion-FAN~\cite{meng_frame_2019} for finetuning.
 To obtain the PCA of the enhanced 3D keypoints, we extract the largest 32 eigenvalue matrix and mean of 2,500 videos from the training set.

 \noindent\textbf{Baselines.} We compare with SOTA one-shot talking-head generation methods on LRW~\cite{chung2016lip} and MEAD~\cite{wang2020mead} test set. They are ATVG~\cite{chen2019hierarchical}, Wav2Lip~\cite{prajwal2020lip}, MakeItTalk~\cite{zhou2020makelttalk}, AVCT~\cite{wang2022one}, PC-AVS~\cite{zhou2021pose} and EAMM~\cite{ji_eamm:_2022}.

\begin{figure}[t!]
  \centering
  \includegraphics[width=0.45\textwidth]{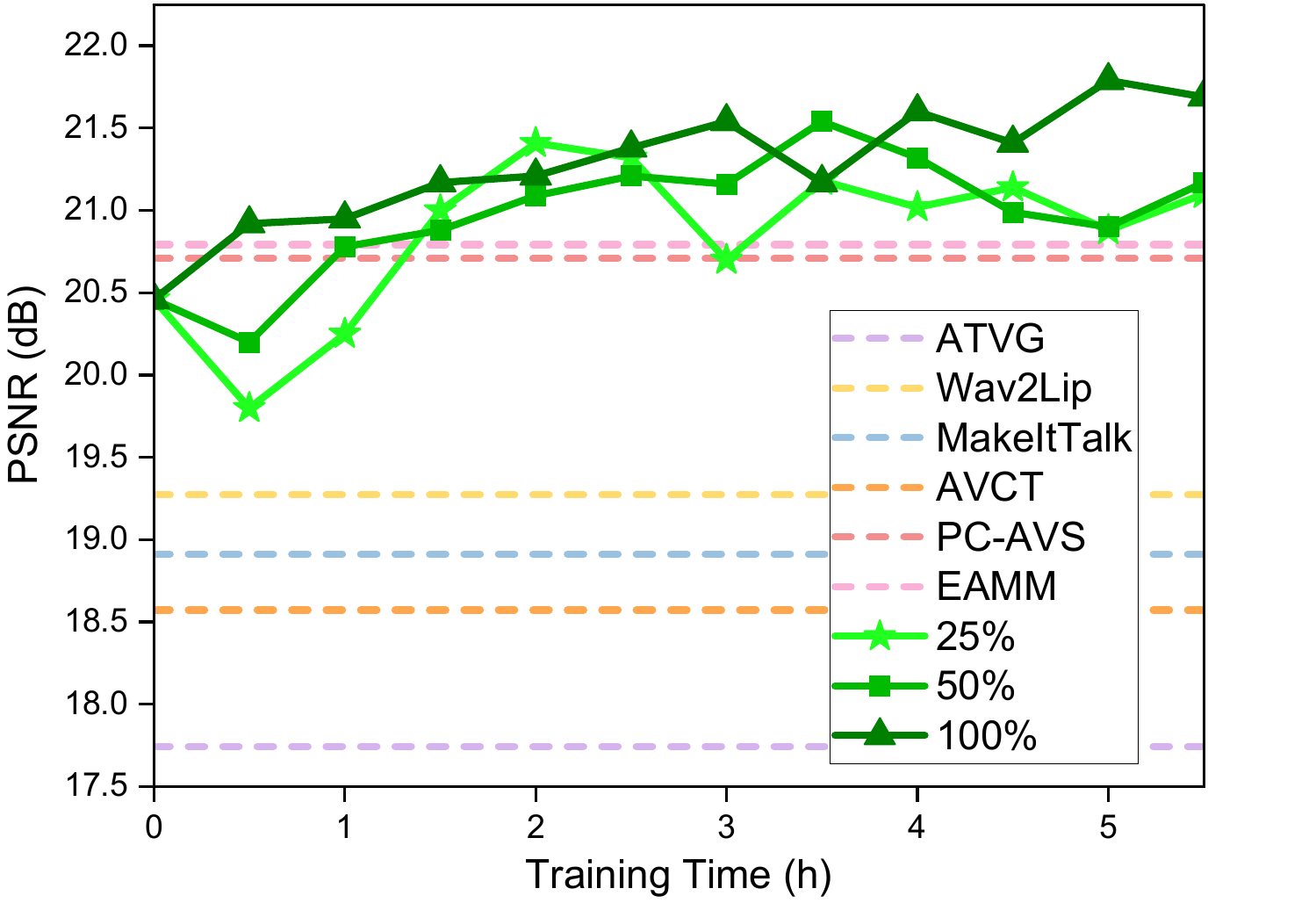}
  \caption{\textbf{Tuning Efficiency.} Our EAT exhibits exceptional tuning efficiency, achieving state-of-the-art performance with just 50\% MEAD training data in a one-hour fine-tuning session, or even with only 25\% data in a two-hour session.
  }
  \label{fig:psnr_time}
\end{figure}

\begin{figure}
  \centering
  \includegraphics[width=0.45\textwidth]{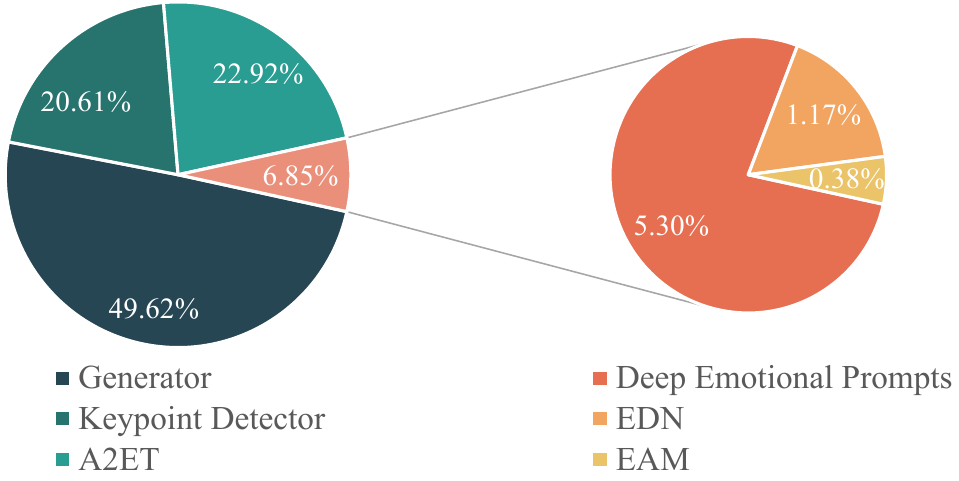}
  \caption{The percentage of parameters in EAT.}
  \label{fig:sup_pie}
  \vspace{-0.2cm}
\end{figure}

\noindent\textbf{Metric.} We assess the quality of synthesized emotional videos with the following metrics: 

\textit{Image quality.} We utilize PSNR, SSIM and Frechet Inception Distance score (FID)~\cite{heusel2017gans} to measure the image quality of synthesized videos.  

\textit{Audio-visual synchronization.} We evaluate the audio-visual synchronization of the synthesized videos using the confidence score of SyncNet ~\cite{chung2016out}. In addition, the distance between the landmarks of the mouth (M-LMD) ~\cite{chen2019hierarchical} is used to indicate speech content consistency, while the distance between the landmarks of the whole face (F-LMD) represents the accuracy of the pose and expressions.

\textit{Emotional accuracy.} To assess the emotional accuracy (Acc$_{emo}$) of the generated emotions, we fine-tune the Emotion-Fan~\cite{meng_frame_2019} using the training set of MEAD.

\subsection{Talking-head Generation}
To verify the effectiveness of EAT, we do experiments on emotion-agnostic and emotional talking-head generation.

\noindent\textbf{Emotion-agnostic Talking-head Generaiton.}
For one-shot emotion-agnostic talking-head generation, we test on the LRW test set, which comprises 25k neutral videos. We take the first frame as the source image for every test video. As presented in Table~\ref{tab:tab_lrw}, our method outperforms other methods in terms of visual quality for emotion-agnostic talking-head generation. Furthermore, our EAT can improve the performance of the pretrained talking-head model. Wav2Lip and PC-AVS risk overfitting the pretrained lip synchronization scoring model since their sync scores surpass the ground truth. Additionally, Wav2Lip only generates mouth parts without facial expression and head pose. 

\begin{figure}[t!]
  \centering
  \includegraphics[width=0.47\textwidth]{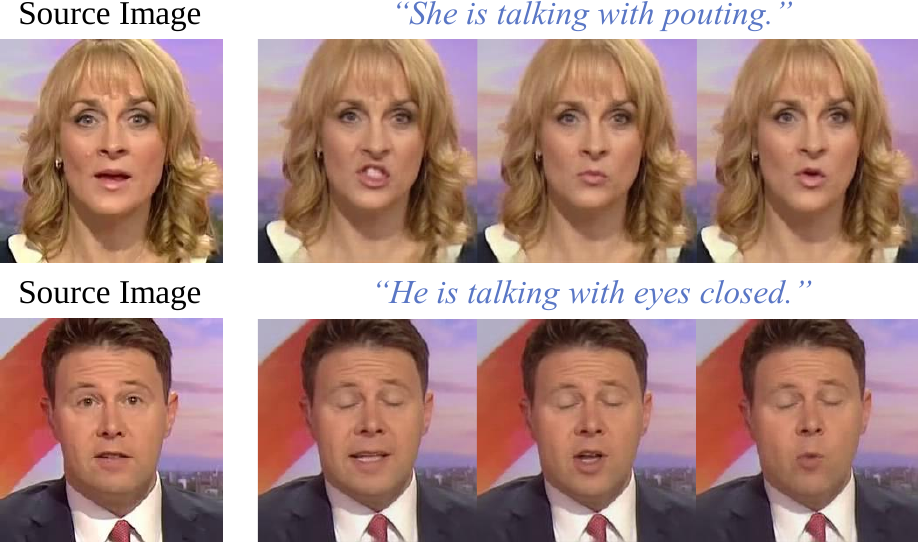}
  \caption{\textbf{CLIP-based zero-shot editing.} The expressions are provided by \textcolor[RGB]{68,114,196}{\textit{``text description"}}. The neutral videos and source images are in-the-wild from LRW~\cite{chung2016lip}.}
  \label{fig:clip}
\end{figure}

\begin{figure}
  \centering
  \includegraphics[width=0.47\textwidth]{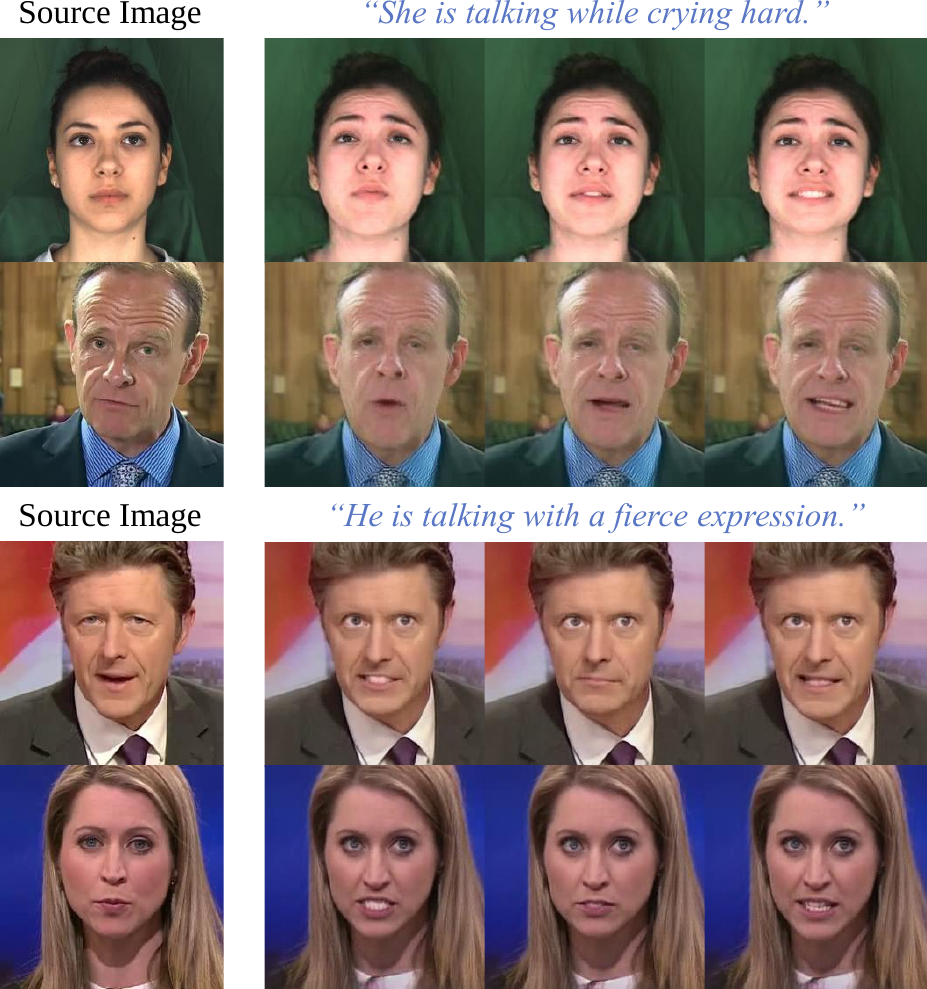}
  \caption{Additional zero-shot results of our EAT. Emotional expressions are given by \textcolor[RGB]{68,114,196}{\textit{"text description"}}. The first row of each text shows the editing results and the second row shows the generated talking head of a different identity with the learned guidance. Please refer to our video for more details. Source images are from MEAD\cite{wang2020mead} and LRW\cite{chung2016lip}.}
  \label{fig:sup_zero}
  \vspace{-0.2cm}
\end{figure}

\begin{table}[t]
  \begin{center}
  \setlength{\tabcolsep}{7pt}
    {
      \small
          \begin{tabular}{l | c c c c}
  \toprule
                  \ Method   &PSNR$\uparrow$   &  M/F-LMD$\downarrow$ &  Sync$\uparrow$ & Acc$_{emo}$$\uparrow$   \\
                  \midrule
              OSFV~\cite{wang2021one}     &22.39  &  1.60/2.12    & 6.68  & 30  \\
              Enhanced &\textbf{24.49}  & \textbf{1.09}/\textbf{1.49}   & \textbf{7.49}  & \textbf{86} \\
  \bottomrule
          \end{tabular}}
  \end{center}
   \caption{\textbf{Ablation study of enhanced latent representation.}  To validate the effectiveness of our enhanced latent representation, we generate emotional videos driven by source images and emotional videos in the MEAD test set.}
  \label{tab:tab_ablation1}
\end{table}

\noindent\textbf{Emotional Talking-head Generation.}
We follow the setting of EAMM to compare the emotional talking-head generation on the public-available MEAD test set. For all methods, the neutral source frames are from EAMM.

Table~\ref{tab:tab_lrw} shows our EAT can achieve the best performance in most metrics. Specifically, EAT achieves better video quality and higher emotion accuracy than other approaches. These findings support the superiority of the emotional representation learned by our proposed method.
Fig.~\ref{fig:qualitative} visually demonstrates our capability to produce authentic and comprehensive emotional expressions in the one-shot setting. Note that AVCT~\cite{wang2022one} is not capable of controlling pose explicitly even with ground-truth 6DoF.
For more results, please refer to our supplementary.

\begin{table}[t]
  \begin{center}
      \setlength{\tabcolsep}{8pt}
      {\small
          \begin{tabular}{l | c c c c}
  \toprule
              Prompt \     &PSNR$\uparrow$   &  M/F-LMD$\downarrow$ &  Sync$\uparrow$ & Acc$_{emo}$$\uparrow$   \\
              \midrule
              w/o          &20.46  & 2.85/2.99   &\textbf{8.12}   & 25   \\
              Shallow      &21.19  &2.50/2.63    & 7.63  & 57  \\
              Deep         &\textbf{21.23}  & \textbf{2.36}/\textbf{2.48}   & 7.83  & \textbf{84} \\
  \bottomrule
          \end{tabular}}
  \end{center}
    \caption{\textbf{Ablation study of prompt.} To verify the effectiveness of shallow and deep emotional prompts, we produce videos using source images and emotional videos from the MEAD test set.}
  \label{tab:prompt}
   \vspace{-0.2cm}
\end{table}

\subsection{User Study}
To evaluate the generated emotional talking heads with in-the-wild images, we conducted a user study with 14 participants to assess lip-sync, video quality, and emotion classification. The maximum score value is 5. To ensure a diverse set of images, we randomly sampled 16 images from both the CREMA-D and LRW datasets, which are not included in the training data. Additionally, we used audio from MEAD test set to generate a total of 32 videos (4 × 8 emotions) for each method. As shown in Table~\ref{tab:userstudy}, our method achieved the best scores for lip-sync, video quality, and emotion accuracy, validating the effectiveness of our EAT architecture in generating emotional talking heads. Besides, the results show that the sync value in Table~\ref{tab:tab_lrw} is inaccurate for emotional talking-head due to the SyncNet model being trained only with neutral talking-head videos. More in-the-wild results can be found in the supplementary.

\subsection{Tuning Efficiency}
During the finetuning of the second stage, we conducted periodic tests on a subset of the MEAD test set every half hour. This enabled us to demonstrate the time and data efficiency of our EAT. EAT can efficiently adapt the pretrained A2ET model to sub-tasks, even with limited data. As shown in Fig.~\ref{fig:psnr_time}, EAT can surpass the SOTA results within one hour with full or half data. We can also achieve comparable performance with just a quarter of the data within two hours. Moreover, as shown in Fig~\ref{fig:sup_pie} the emotional adaptation modules require only 6.85\% additional parameters compared to the pretrained model. Deep emotional prompts account for 5.30\%, EDN for 1.17\%, and EAM for 0.38\% of the additional parameters. These results demonstrate the effectiveness and efficiency of our EAT approach.

\begin{table}[t]
    \setlength{\tabcolsep}{6pt}
    \begin{center}
        {\small
            \begin{tabular}{l | c c c c}
                \toprule
                Ablation & (A)   & (B)         &  (C) &\textbf{EAT}       \\
                \midrule
                Prompt   &       &   \checkmark & \checkmark   & \checkmark   \\
                EDN      &       &     & \checkmark  & \checkmark           \\
                EAM      &       &     &    & \checkmark            \\
                \midrule
                PSNR$\uparrow$     & 20.46  & 21.23  &  21.40 & \textbf{21.79} \\
                M/F-LMD$\downarrow$  & 2.85/2.99   & 2.36/2.48   &  2.28/\textbf{2.41}  & \textbf{2.22}/2.43 \\
                Sync$\uparrow$     & 8.12   & 7.83   &  7.83  & \textbf{8.22}\\
                Acc$_{emo}$$\uparrow$      & 25     &  \textbf{84}    & 81     &  67   \\ 
  \bottomrule
            \end{tabular}}
    \end{center}
        \caption{\textbf{Ablation study of each component.} Each component contributes to the improvement of the video quality, thus verifying its effectiveness.}
    \label{tab:tab_ablation2}
\end{table}

\begin{table}[t]
    \setlength{\tabcolsep}{6.5pt}
    \begin{center}
        {
            \begin{tabular}{l | c c c c}
               
                \toprule
                $\mathcal{L}_{per}$   &  \checkmark &   \checkmark & \checkmark   & \checkmark   \\
                $\mathcal{L}_{lat}$   &       & \checkmark & \checkmark  & \checkmark           \\
                $\mathcal{L}_{sync}$  &       &   &  \checkmark  & \checkmark            \\
                $\mathcal{L}_{1}$     &       &   &    & \checkmark            \\
                \midrule
                PSNR$\uparrow$     & 21.52  & 21.61  &  21.31 & \textbf{21.79} \\
                Sync$\uparrow$     & 5.50   & 5.66   &  8.13  & \textbf{8.22}\\
  \bottomrule
            \end{tabular}}
    \end{center}
        \caption{\textbf{Ablation study of each loss.} Each loss contributes to the improvement of the video quality or synchronization value.}
    \vspace{-0.3cm}
    \label{tab:tab_ablationloss}
\end{table}

\subsection{Zero-shot Expression Editing}
We conduct zero-shot expression editing with the CLIP ~\cite{radford2021learning} model to generate a novel emotional talking head video, as shown in Fig.~\ref{fig:clip}. Given a neutral video, we treat the first frame as the source image and edit the expression by text descriptions. Based on EAT architecture, we learn the emotional guidance and EAM with an additional CLIP loss~\cite{radford2021learning}. We notice that text descriptions will determine the editing performance and need careful design. Besides, the learned latent code from one video and guidance texts can also be applied to another video. As shown in Fig.~\ref{fig:sup_zero}, We present different identity results manipulated by a learned latent code.

\subsection{Ablation Study}

To assess the effectiveness and importance of various aspects of EAT, We conduct several ablation studies on our proposed architecture and modules. 

{\noindent\bf{Enhanced Latent Representation. }}We compare the latent representation before and after our proposed enhancement. As shown in Table~\ref{tab:tab_ablation1}, the superior performance in face reenactment shows that our pre-trained model can capture a wider range of emotional facial movements than OSFV~\cite{wang2021one}, which has been discussed in Sec.~\ref{sec:3.1.1}.

{\noindent\bf{Prompt. }}To study the effect of different types of prompts, we conduct experiments with shallow and deep emotional prompts. Table~\ref{tab:prompt} shows that deep prompt learns emotional expression deformation better than shallow prompt, although it has a side effect on synchronization.

{\noindent\bf{Each Component. }}To verify the effectiveness of our proposed modules, we conduct ablation experiments by removing one component every time. Table~\ref{tab:tab_ablation2} shows that all three components can improve video quality. Deep emotional prompts can transfer the talking-head knowledge to emotional talking-head generation at the expense of synchronization. Although deep emotional prompts lead to intense emotional expressions, the outputs deviate from the ground truth. By incorporating EDM and EAM, the image quality and fidelity toward the ground truth are enhanced, as evidenced by the rise in PSNR/SSIM values. Yet, this comes at the expense of emotion intensity and accuracy. Please refer to our supplementary for visual analysis.

{\noindent\bf{Each Loss. }}As shown in Table~\ref{tab:tab_ablationloss}, we conduct an ablation study on perceptual loss, latent loss, sync loss, and $\mathcal{L}_1$ loss in the finetuning stage. It demonstrates that sync loss contributes to synchronization while others contribute to expression fidelity.

\section{Conclusion}
In this paper, we propose an efficient emotional adaptation paradigm for audio-driven talking-head generation, consisting of two stages. First, we enhance 3D latent representation and develop a transformer architecture A2ET to achieve emotion-agnostic talking-head generation. Second, we introduce learnable guidance for emotional expression control through our deep emotional prompts, EDN, and EAM. With these adaptation modules, EAT can quickly transfer the pretrained talking-head model to emotional talking-head generation. Experiments demonstrate that our EAT is the first parameter-efficient and effective paradigm for emotional talking-head generation. 

\noindent\textbf{Limitations and Broader Impact.} 1) The drawbacks of emotional training data, such as the diversity of background and head poses, will affect the generalizability of our EAT. 2) Our approach paves the way for broader talking-head applications, including zero-shot or one-shot emotional talking-head generation.

\noindent\textbf{Acknowledgements.} This work is supported by the National Key R\&D Program of China (2022YFB3303300) and the Fundamental Research Funds for the Central Universities (No. 226-2023-00048).


{\small
\bibliographystyle{ieee_fullname}
\bibliography{eg}
}


\newcommand{\appendixhead}
{\centering{\Large \bf Supplementary Material}
\vspace{10mm}}

\twocolumn[\appendixhead]

\appendix
\section{The Networks Details}  
We provide additional details of our network architecture and training procedure. It should be noted that the Keypoint Detector ($D_{k}$) and RePos-Net networks are primarily derived from OSFV \cite{wang2021one}. For more information, interested readers may refer to OSFV \cite{wang2021one}.

\begin{figure}
  \centering
  \includegraphics[width=0.48\textwidth]{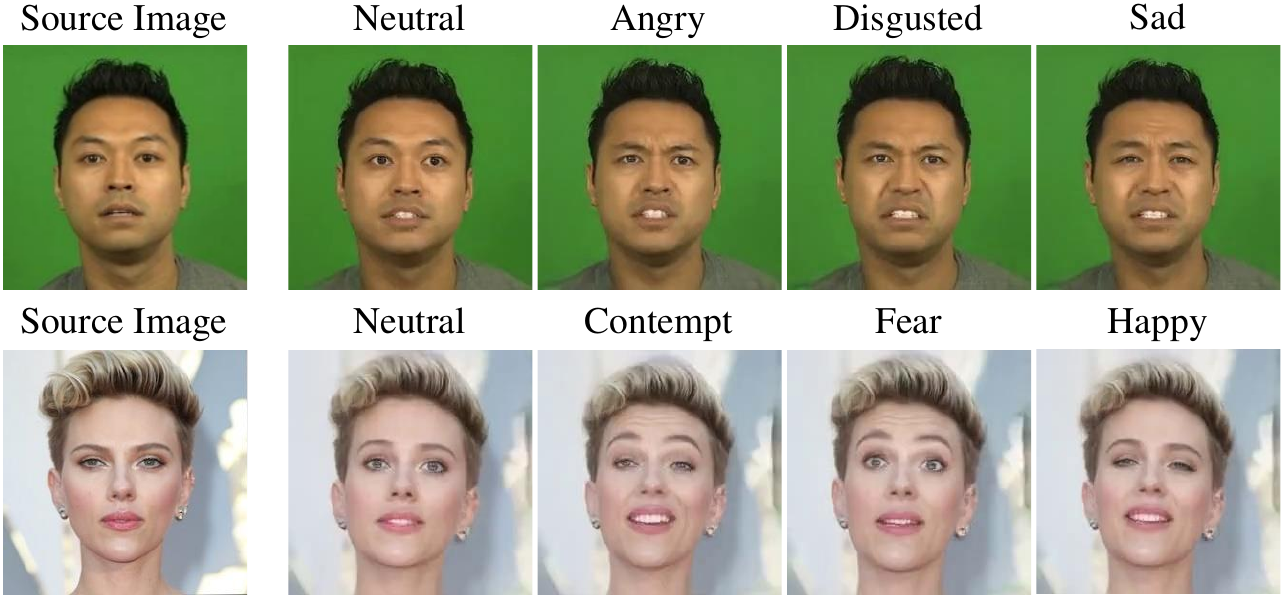}
  \caption{Additional emotional expressions generated by EAT. EAT produces realistic and diverse facial expressions with corresponding emotional guidance. Please zoom in for a better view. Source images are from CREMA-D\cite{cao2014crema} and MakeItTalk\cite{zhou2020makelttalk}.}
  \label{fig:sup_exp}
\end{figure}

\begin{figure*}
  \centering
  \includegraphics[width=1.0\textwidth]{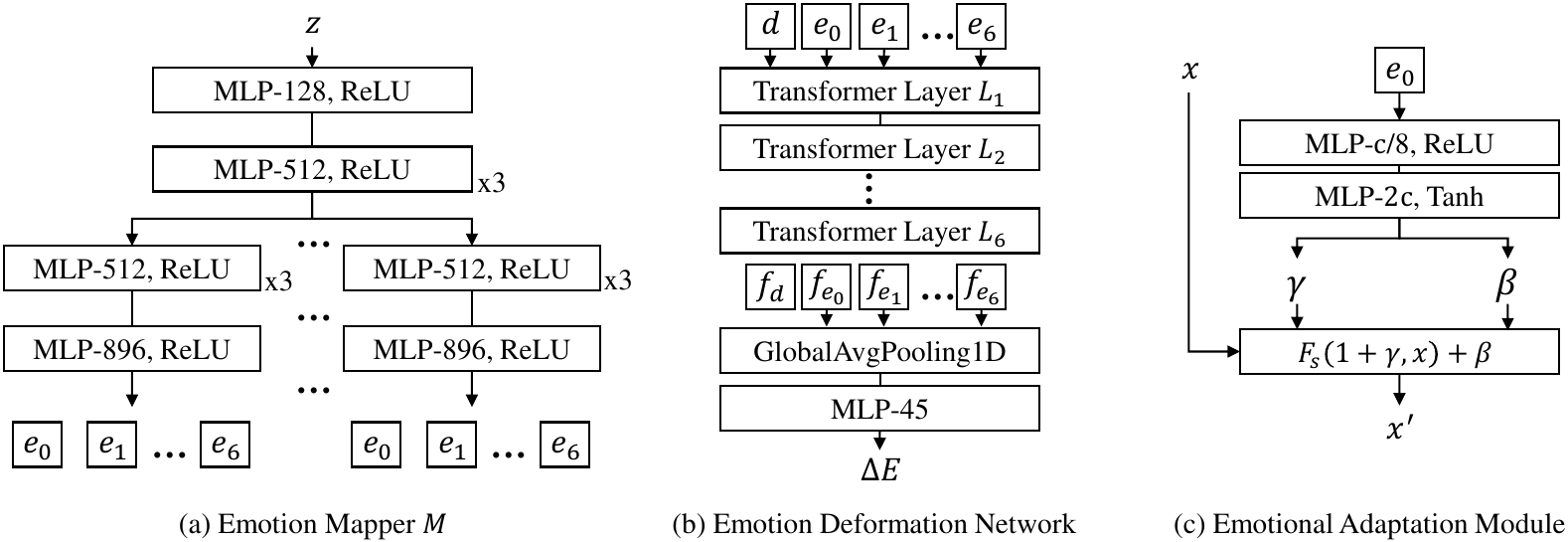}
  \caption{More network architectures of our EAT model.}
  \label{fig:sup_arch}
\end{figure*}

\begin{table*}[h]
\begin{center}
\scalebox{1.0}{
\begin{tabular}{l|cccccccccccc}
\toprule

                & Happy & Angry & Disgusted & Fear  & Sad  & Neutral  & Surprised  & Contempt & Average\\ 

\midrule
Wav2Lip~\cite{prajwal2020lip} & 0.00 & 25.64 & 0.00 & 0.00 & 0.00 & 91.25 & 0.00 & 0.00 & 17.87 & \\
MakeItTalk~\cite{zhou2020makelttalk} & 0.00 & 25.64 & 0.00 & 0.00 & 0.00 & 75.00 & 0.00 & 0.00 & 15.23 & \\
AVCT~\cite{wang2022one}             & 0.83 & 25.64 & 0.00 & 0.00 & 0.00 & 69.38  & 0.00 & 10.08 & 15.64 & \\
EAMM~\cite{ji_eamm:_2022}            & 23.33 & 84.48 & 9.40 & 0.00 & 0.00 & 98.13 & 94.02 & 72.27 & 49.85 & \\
\midrule
Pretrain (Ours)  & 35.00 & 11.97  & 0.00 & 0.00  & \textbf{49.17}  & 38.75  & 0.00 & 59.66 & 25.18 &\\
EAT (Ours)       & \textbf{84.17} & \textbf{100.00} & \textbf{48.72} & \textbf{16.52} & \textbf{49.17} & \textbf{100.00} & \textbf{100.00} & \textbf{94.96} & \textbf{75.43}\\
\bottomrule
\end{tabular}
}

\end{center}
\caption{Quantitative evaluation of the emotion classification in the MEAD dataset.}
\label{tab:tab_emo}
\vspace{-0.3cm}
\end{table*}

{\noindent \bf Audio-to-Expression Transformer.} We use the Audio-to-Expression Transformer (A2ET) to transfer the audio to 3D latent expression deformation sequences. The A2ET consists of an encoder and a decoder, both with 6 transformer layers and 8 heads. The feed-forward layer has a dimension of 1024. Each token is a 128-dim vector. The expression deformation vector ($E_{i}$) is predicted by the feature of the central frame $i$. However, directly optimizing the 3D expression motions leads to convergence problems in network training. To address this issue and bridge the gap between the 3D expression deformation and the audio features, we use principal component analysis (PCA) to reduce the dimensionality of $E_{i}$ from 45 to 32. Specifically, we calculate the matrix of principal eigenvalues $U$ and mean vector $M$ from the training set. Then the expression deformation vector is obtained by projecting the predicted PCA using the following equation:
\begin{equation}
E_{i} = PE_{i} * U^{T} + M, \label{eq:pca2lmk}
\end{equation}
where $PE_{i}$ is the predicted PCA and $E_{i}$ is the expression deformation, which is used to modify the neutral 3D keypoints to generate the expressive face. As the number of keypoints is 15, the shape of $E_{i}$ is (15, 3).

{\noindent \bf Emotion Mapper.} 
We propose an emotion mapper that produces emotional tokens to guide the generation of emotional expressions. As shown in Fig.~\ref{fig:sup_arch}(a), 
the emotion mapper $M$ consists of several shared and unshared multi-layer perceptrons (MLP) layers. 
It takes a 16-dim latent code $z$ as input and outputs seven emotional tokens $e_{0}$, $e_{1}$, $\cdots$, $e_{6}$. 
The first token $e_{0}$ serves as the emotional guidance for the emotional adaptation module (EAM), which modifies the features of the audio-to-expression transformer (A2ET). 
The remaining six tokens $e_{1}$, $\cdots$, $e_{6}$ are fed to the corresponding transformer layer of A2ET as deep emotional prompts. 
The Emotional Deformation Network (EDN) then uses all these tokens and the latent source representation to generate the emotional deformation $\Delta E$.

{\noindent \bf Emotional Deformation Network.} The Emotional Deformation Network (EDN) learns the emotional deformation $\Delta E$ using the same architecture as the A2ET encoder, which has six transformer layers. 
Fig.~\ref{fig:sup_arch}(b) shows the input and output of EDN. It takes the latent source representation $d$ and the emotional guidance tokens $e_{0}$, $e_{1}$, $\cdots$, $e_{6}$ as input, and extracts their features $f_{d}$, $f_{e_{0}}$, $\cdots$, $f_{e_{6}}$. 
Then it applies global average pooling to the emotion-related features $f_{e_{0}}$, $\cdots$, $f_{e_{6}}$ and uses an MLP layer to obtain the final emotional deformation $\Delta E$.

{\noindent \bf Emotional Adaptation Module.}  The emotional adaptation module (EAM) consists of two multi-layer perceptrons (MLPs). As shown in Fig.~\ref{fig:sup_arch}(c), given the input feature $x$ and the emotional token $e_0$, we extract the weight vector $\gamma$ and the bias vector $\beta$ using MLPs. They have the same dimension as the input $x$. With the channel-wise multiplication operation $F_s$ and channel-wise addition, we obtain the output $x'$.

\begin{table}[t]

  \begin{center}
      \scalebox{1.0}{
          \begin{tabular}{l | c c c c}
\toprule
              Weight \     &PSNR$\uparrow$   &  M/F-LMD$\downarrow$ &  Sync$\uparrow$ & Acc$_{emo}$$\uparrow$   \\\midrule
              w/o         &21.49  & 2.27/2.46   &8.02   & \textbf{76}   \\
              EAT         &\textbf{21.79}  & \textbf{2.22}/\textbf{2.43}   & \textbf{8.22}  & 67 \\
              \bottomrule
          \end{tabular}}
  \end{center}
    \caption{\textbf{Ablation study of EDN weight initialization.} The weight initialization of EDN with the A2ET encoder promotes the performance of EAT.}
    \label{tab:weightinit}
\end{table}

\begin{figure*}
  \centering
  \includegraphics[width=1.0\textwidth]{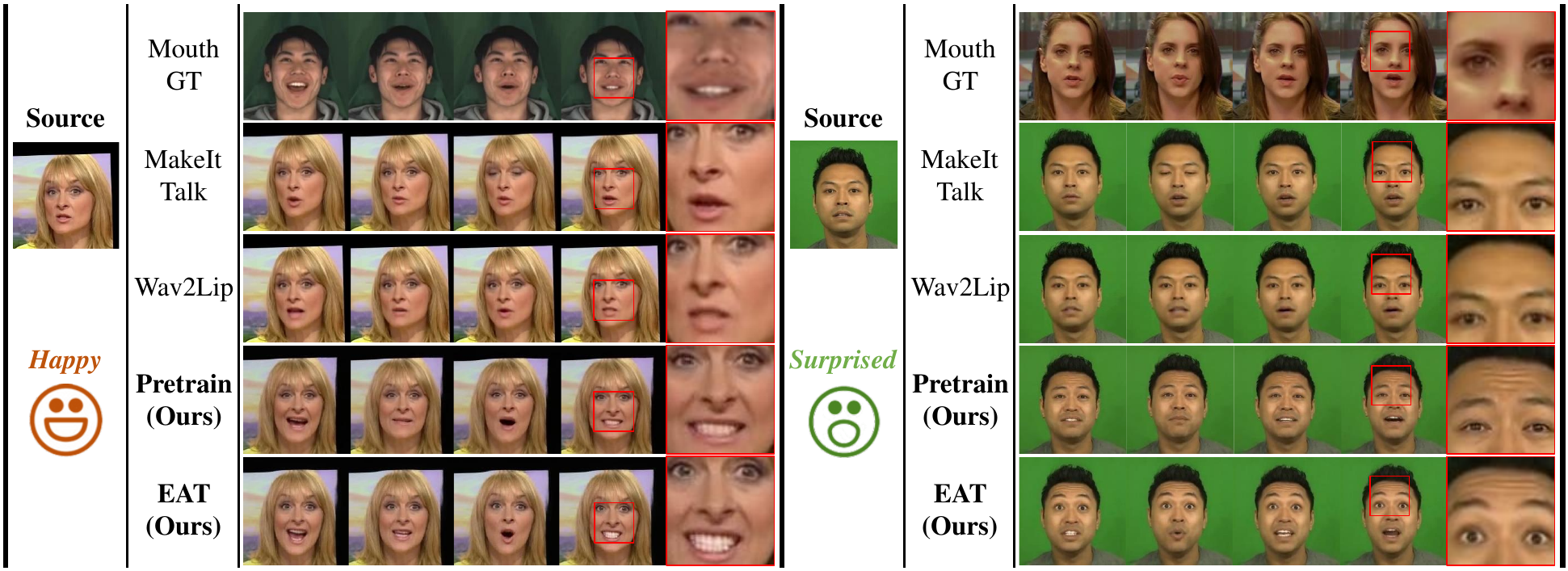}
  \caption{More Qualitative results. We compare with more baselines, such as MakeItTalk~\cite{zhou2020makelttalk}, Wav2Lip~\cite{prajwal2020lip}, and our pretrained model.}
  \label{fig:sup_morebaseline}
\end{figure*}

\begin{figure*}
  \centering
  \includegraphics[width=1.0\textwidth]{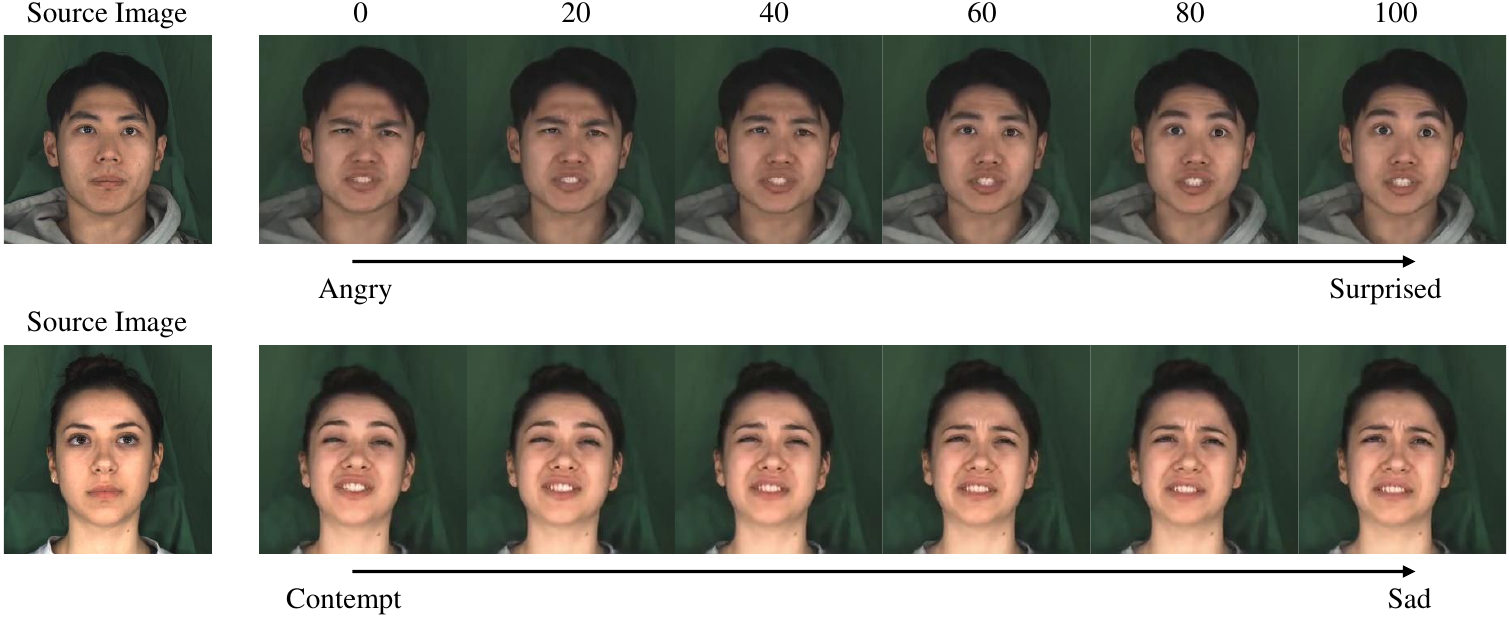}
  \caption{Emotion interpolation. The top row is the emotion interpolation results between $Angry$ and $Surprised$. The bottom row is the results between $Contempt$ and $Sad$. The neutral faces are from MEAD~\cite{wang2020mead}}
  \label{fig:sup_emointer}
  \vspace{-0.3cm}
\end{figure*}

{\noindent \bf Parameter Efficiency Analysis.} Our Deep Emotional Prompts, EDN and EAM only require about 7\% of the parameters compared to the whole network. The emotion mapper, which generates deep emotional prompts for eight emotions, has most of the parameters. In addition, EDN and EAM consume less than 2\%. These parameters are 13.9M. This is half of the emotional network of EAMM~\cite{ji_eamm:_2022}, which has 27.9M parameters.

\section{Training and Testing Details}  
{\noindent \bf Training Details.}
We use the MEAD dataset and 8k emotional video clips from Voxceleb2~\cite{chung_voxceleb2:_2018} with various facial expressions to learn the enhanced latent keypoints. We also use roughly 21k emotional images from AffectNet~\cite{8013713} to train emotional expression generation. Due to the lack of corresponding neutral faces, we generate neutral faces paired with emotional images by using Ganimation~\cite{pumarola2018ganimation}.
We train our EAT with Adam~\cite{kingma2014adam} with $\beta_{1}=0.5$ and  $\beta_{2}=0.999$. The learning rate is set to $1.5\times 10^{-4}$ for A2ET and $2\times 10^{-4}$ for other modules. In the first stage, we train A2ET with only latent loss first to obtain a good initialization, and then we train it with full loss. To improve generalization, we use the Voxceleb2 and MEAD datasets, which contain about 225k video clips. In the second stage, we finetune efficient adaptation modules with only the MEAD dataset, which has about 10k video clips. We test our model on LRW~\cite{chung2016lip} and MEAD~\cite{wang2020mead} dataset.

{\noindent \bf Testing Details and Protocol.}
When testing LRW, the input is the first frame, and the transformation starts from the first frame. Therefore, the relative offsets of the latent keypoints are used. When testing MEAD, due to the variation in facial expressions, which is unrelated to the neutral source image, the predicted latent keypoints are used. 

To ensure accurate evaluations, we crop and align ~\cite{chen2019hierarchical} the faces before calculating these metrics: PSNR, SSIM, FID, M-LMD, and F-LMD. As for synchronization confidence, we preprocess the generated videos with reference to PC-AVS ~\cite{zhou2021pose}. 

\begin{figure*}[h]
\begin{center}
   \includegraphics[width=0.8\linewidth]{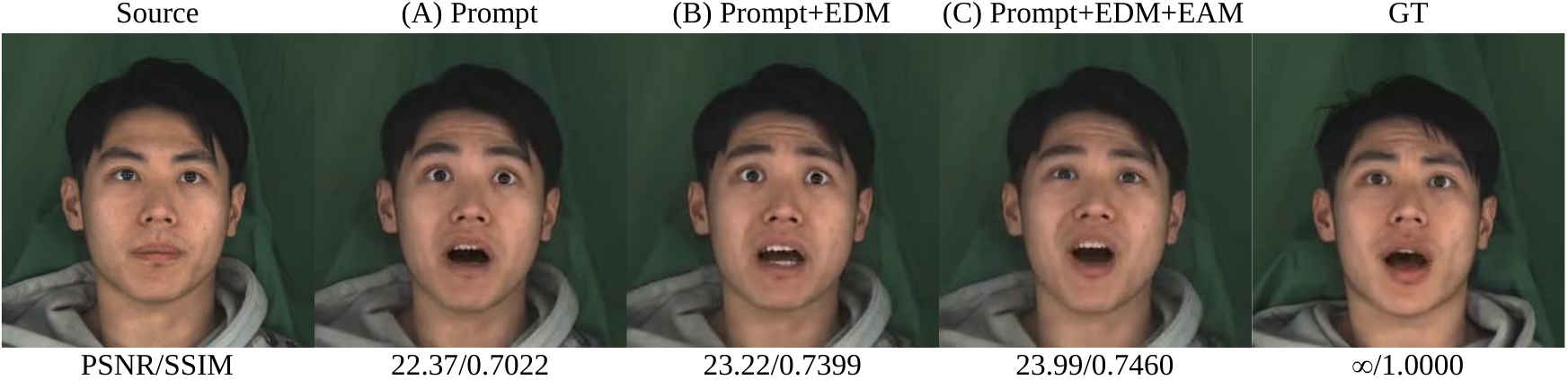}
\end{center}
   \vspace{-.3cm}
   \caption{Visualization on each component of EAT.}
  \label{fig:sup_visualeachcomp}
\end{figure*}

\begin{figure}
\begin{center}
   \includegraphics[width=1\linewidth]{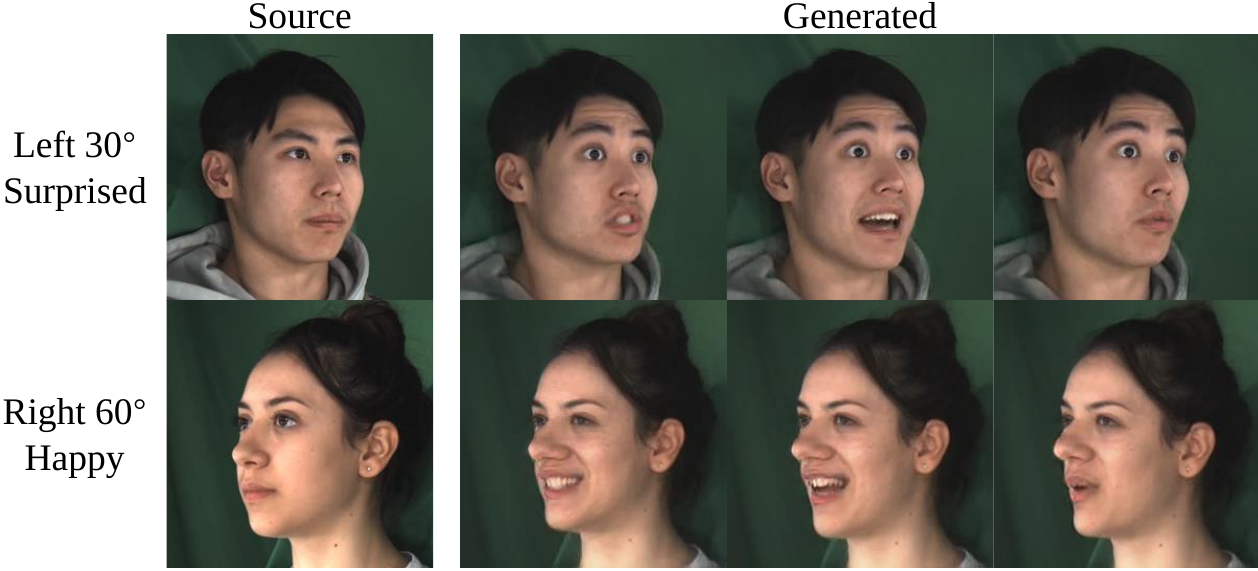}
\end{center}
   \vspace{-.3cm}
   \caption{Visualization on the profile faces of MEAD.}
  \label{fig:sup_profile}
\end{figure}

\begin{figure}
  \centering
  \includegraphics[width=0.48\textwidth]{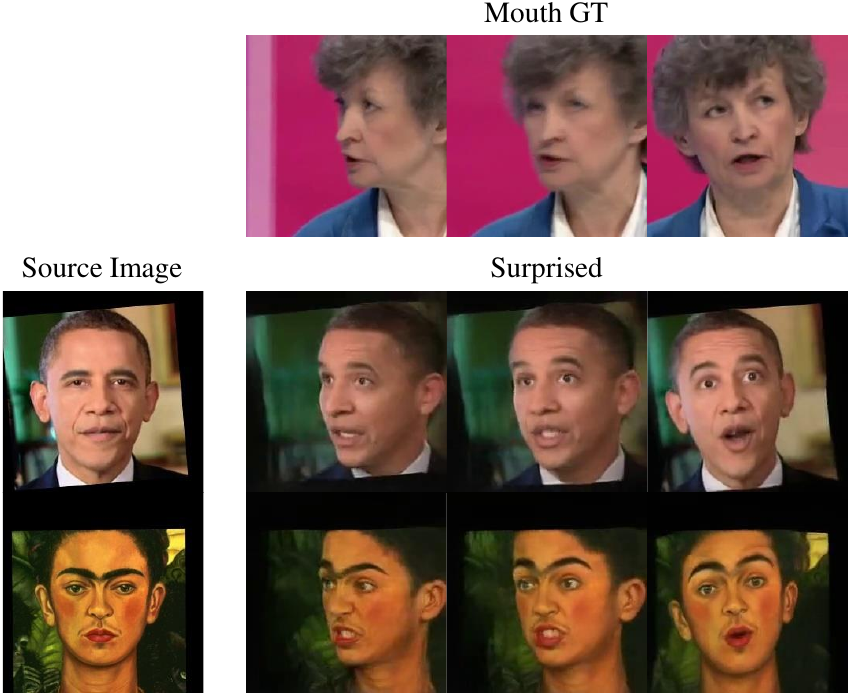}
  \caption{Failure case. The driving audio and poses are from the videos in the first row. The second and third rows display the generated results with $Surprised$ emotional guidance. Neutral faces are from MakeItTalk~\cite{zhou2020makelttalk} and driving video is from LRW~\cite{chung2016lip}.}
  \label{fig:sup_fail}
\end{figure}

\section{Additional Experimental Results}

{\noindent \bf Additional baseline results.} As shown in Figure~\ref{fig:sup_morebaseline}, we compare our EAT results with several baseline methods. Our results are more pleasant than those of MakeItTalk~\cite{zhou2020makelttalk} and Wav2Lip~\cite{prajwal2020lip}, which do not consider emotional expression in talking heads. Additionally, our EAT achieves emotion control compared to the pretrained A2ET network. Videos are included in the supplementary material for reference.

{\noindent \bf Various emotional expressions} To validate the diversity of emotional expressions generated by EAT, we present six different emotional results in Fig.~\ref{fig:sup_exp}. Compared to $Neutral$ emotion, emotional expressions result in different modifications to facial elements, such as lip corners, eyes, and brows. We present the quantitative results of emotion classification in Table~\ref{tab:tab_emo}. We notice that EAT works significantly better on $Happy$, $Sad$, $Disgusted$, and $Contempt$ than other methods. This is because our method can capture mouth details and these emotions can be more clearly reflected by the lips. As for $Neutral$, $Angry$, and $Surprised$, EAMM~\cite{ji_eamm:_2022} performs well since these emotions are more apparent on the eyes and brows. And EAT can also achieve better performance in these emotions. However, all methods perform poorly on $Fear$ emotion. It may be because $Fear$ and $Surprise$ are similar and difficult to distinguish.

{\noindent \bf Emotion interpolation.} We conduct emotional guidance interpolation on the MEAD test set to verify that the latent space learned by the emotion mapper is continuous, as Fig.~\ref{fig:sup_emointer} shows.

{\noindent \bf Additional ablation study.} 
We conduct further ablation studies on the weight initialization of EDN, Our results, presented in Table~\ref{tab:weightinit}, show that using the weight initialization of the A2ET encoder leads to quicker convergence and improved performance in terms of video quality and audio-visual synchronization.

{\noindent \bf Visual analysis on each component of EAT.} To analyze the effect of each component of our model, we show the fear emotion
results from (A), (B), and (C), with corresponding accuracy rates of 38.46\%, 30.77\%, and 15.38\% respectively in Fig.~\ref{fig:sup_visualeachcomp}. Deep emotional prompts help generate intense emotional expressions that deviate from the Ground Truth (GT). By incorporating EDM and EAM, we achieve greater fidelity toward the GT and higher image quality in terms of PSNR/SSIM. This is attributed to the learning capabilities of EDM and EAM for emotional data. However, it results in reduced emotion intensity and accuracy.

{\noindent \bf Visualization on the profile faces.} To assess the ability of enhanced latent representation in 3D talking-head generation, as shown in Fig.~\ref{fig:sup_profile}, we visualize the talking-head frames generated from the profile faces of MEAD. We test the faces captured from left 30 degrees and right 60 degrees with $Suprised$ and $Happy$ emotions.

\section{Limitations and Future Work.}
While EAT is capable of generating emotional talking-head videos with emotional guidance, there are still some limitations. Firstly, the diversity of background and head pose in emotional training data can affect the generalizability of our EAT. As shown in Fig.~\ref{fig:sup_fail}, the wrinkles on the forehead are not obvious in these in-the-wild images. This issue could be addressed by more naturalistic and non-acted emotional data~\cite{zafeiriou2017aff, kossaifi2019sewa, busso2008iemocap} and representations with the head prior, such as FLAME\cite{zheng2022avatar}. Secondly, effective guidance texts are required to achieve zero-shot generation. This may be due to the limited ability of models trained on image-text pairs to capture emotional expression, which could affect the performance of zero-shot learning. Thirdly, the eye region, such as eye blinks~\cite{vougioukas2020realistic} and gaze direction~\cite{doukas2023free}, has not been considered in our work. Finally, the discrete emotion guidance limits the representation ability of our model. It needs to note that facial expressions are not always representative of the internal emotional state~\cite{barrett2019emotional}. More refined theories of emotion, such as the valence-arousal model, may help generate more realistic emotions. We leave these problems for future work.

\section{Ethical Considerations.}
Our research is intended for use in virtual human research and entertainment. However, there is a risk that the emotional talking-head generation algorithm could be abused. We strongly recommend that generated talking-head videos be labeled as “fake”. On one hand, our work demonstrates that emotional talking-head generation is technically feasible. On the other hand, fake video detection~\cite{cozzolino2021id, masood2021deepfakes} has attracted significant attention. We would be happy to assist in the development of related research.

\ificcvfinal\thispagestyle{empty}\fi

\end{document}